\documentclass[graybox]{SNmult}

\usepackage{type1cm}        
\usepackage{makeidx}         
\usepackage{graphicx}        
\usepackage{multicol}        
\usepackage[bottom]{footmisc}

\usepackage{newtxtext}       %
\usepackage[varvw]{newtxmath}       
\usepackage{xcolor} 

\usepackage{microtype}   
\usepackage{float}       

\usepackage{subcaption}
\usepackage[authoryear,round]{natbib}

\usepackage{pgfplots}
\pgfplotsset{compat=1.18}

\usepackage{multirow} 
\usepackage[most]{tcolorbox}

\usepackage[normalem]{ulem}

\makeindex             

\begin{document}

\title*{Scientific Knowledge Discovery in the Age of Large Language Models}
\author{Eleni Adamidi\orcidID{0000-0001-9925-1560}, Serafeim Chatzopoulos\orcidID{0000-0003-1714-5225} and\\Thanasis Vergoulis\orcidID{0000-0003-0555-4128}}
\institute{Eleni Adamidi \at IMSI, ATHENA RC, \email{eleni.adamidi@athenarc.gr}
\and Serafeim Chatzopoulos \at IMSI, ATHENA RC, \email{schatz@athenarc.gr}
\and Thanasis Vergoulis \at IMSI, ATHENA RC, \email{vergoulis@athenarc.gr}}
%
%
\maketitle
\vspace{-2.5cm}
\abstract 
{The rapid growth of scholarly literature has made identifying relevant publications increasingly difficult, and conventional search systems still depend heavily on manually formulated queries and effortful manual inspection. Generative large language models (LLMs) offer a more flexible alternative, supporting literature retrieval and the screening of candidate studies against eligibility criteria. This chapter surveys $34$ peer-reviewed papers applying generative LLMs to these two tasks, identified via a Boolean search over the OpenAIRE Graph ($1,589$ records screened to $34$ inclusions). Reviewed studies are characterised by LLMs employed, model access and adaptation, prompting and architectural techniques, ground-truth sources, and evaluation metrics.}


\section{Introduction}
\label{sec:introduction}

The volume of scientific literature continues to expand rapidly, complicating researchers' efforts to identify, assess, and keep pace with relevant advances in their fields \citep{Hanson2024}. This growth shows no sign of slowing and may be accelerating further with AI writing assistants that enable faster article drafting and publication. At this scale, researchers risk overlooking relevant work, duplicating existing efforts, or incorporating evidence after substantial delay. 
These pressures intensify the need for more effective methods to support scientific knowledge discovery.

This chapter surveys peer-reviewed work on the use of generative \emph{large language models} (LLMs)\footnote{From now on, we use the term LLM specifically for generative LLMs.} to support two core scientific knowledge-discovery tasks: \emph{literature retrieval}, which concerns identifying, ranking, or recommending publications relevant to a query, seed paper, user profile, or research context, and \emph{literature screening}, which concerns assessing candidate publications against predefined eligibility criteria, typically in systematic reviews and other evidence-synthesis workflows.

Traditionally, retrieval has relied on keyword matching \citep{GusenbauerHaddaway2020}, citation-based ranking \citep{Bascur2023}, and recommender techniques such as content-based and collaborative filtering \citep{Beel2016}, while screening has depended largely on manual review, with semi-automated tools (e.g., Abstrackr, EPPI-Reviewer) mainly helping to prioritize or classify records for human assessment \citep{Marshall2019}. Despite these advances, both tasks remain labor-intensive. LLMs offer a more flexible alternative since they can interpret natural-language requests, capture semantic relationships, and produce structured judgments or explanations \citep{Brown2020,Ouyang2022}. The chapter reviews these approaches 
with the goal of providing a comprehensive overview of the landscape and identifying recurring patterns.

\section{Survey Methodology and Literature Overview}
\label{sec:corpus}

\subsection{Search and Screening Strategy}
\label{sec:corpus:search}

Our survey was motivated by the need to characterize the rapidly evolving landscape of LLM-based methods for scientific literature retrieval and screening, aiming to assemble a corpus broad enough to cover the main methodological paradigms and application domains, yet focused enough to support meaningful synthesis.
Relevant literature was identified through BIP!~Finder~\citep{Vergoulis2019},\footnote{BIP! Finder: \url{https://bip.athenarc.gr/}} a scholarly search engine indexing $250$+ million research publications using the OpenAIRE Graph~\citep{manghi_paolo_2022_7488618},\footnote{OpenAIRE Graph: \url{https://graph.openaire.eu/}} as its primary data source.
The search was restricted to publications from 2019 onward, covering the period in which LLMs became practically available in research-support contexts.
The following Boolean query was applied:

\begin{tcolorbox}[title={Search Query},colback=gray!5,colframe=gray!50]
\ttfamily\small
("large language model" OR LLM OR agentic) AND

("literature exploration" OR "scientific exploration" OR
"scientific knowledge discovery" OR "literature discovery" OR
"academic search" OR "scholarly search" OR
"paper recommendation" OR "citation recommendation" OR
"study recommendation" OR "literature retrieval" OR
"paper retrieval" OR "publication retrieval" OR
"literature screening")
\end{tcolorbox}

The query was designed to capture work that explicitly addresses LLM-based methods for knowledge discovery tasks, with the focus on retrieval and screening. The search was conducted in May 2026; hence, the landscape presented in this survey reflects the state of the literature available at that time, based on BIP! DB-v20.1\footnote{BIP! DB is the underlying database powering BIP! Finder.}~\citep{vergoulis2021bip}.

The search query returned $1,589$ candidate papers, which were subsequently deduplicated and screened for eligibility according to predefined inclusion and exclusion criteria (see below). Screening was performed in two stages: an initial assessment based on the title and abstract, followed by a full-text review for papers that met the initial screening criteria. Titles and abstracts were collected automatically by BIP! Finder, whereas full-text articles were manually retrieved and assessed by the authors, who served as domain experts throughout the screening process. 

To be included, a study had to have undergone peer review and satisfy all of the following criteria: (i) it addressed at least one of the target tasks; (ii) it used a generative large language model as a primary component of the proposed system or method; (iii) it reported a system, method, or evaluation with sufficient methodological detail to support comparison with related work; and (iv) the manuscript is written in English.

Studies were excluded if they met any of the following criteria: (i) they were extended abstracts lacking sufficient methodological detail; (ii) they had not undergone peer review\footnote{For each preprint identified during screening, the authors manually checked for a peer-reviewed version not yet indexed by the search engine before final exclusion.} (e.g., preprints, technical reports); (iii) the LLMs addressed tasks outside the scope of this survey (e.g., publication summarisation); (iv) the manuscript was not written in English; or (v) the full text could not be obtained.

\subsection{The Final Corpus}
\label{sec:corpus:stats}

After deduplication, title-and-abstract screening, and full-text review, the final corpus comprised 34 papers published between 2024 and 2026. The temporal distribution (Figure~\ref{fig:temporal}, left) shows rapid growth, from $9$ papers in 2024 to $24$ in 2025, with $1$ additional paper from early 2026 captured at the time of data collection. This pattern suggests the rapid uptake of LLMs in research workflows (the 2026 count is necessarily incomplete because the search was conducted early in this year).
Citation patterns (Figure~\ref{fig:temporal}, right) reflect both recency and early-mover advantage. The 2024 papers have accumulated the most citations relative to their number, while the much larger 2025 cohort has received fewer citations so far, consistent with the citation lag phenomenon~\citep{smith200930}.

\begin{figure}[t]
\centering
\begin{minipage}[t]{0.46\textwidth}
  \centering
%

\begin{tikzpicture}
\begin{axis}[
    ybar,
    bar width=24pt,
    width=6cm,
    height=4.5cm,
    enlarge x limits=0.45,
    xlabel={Year},
    ylabel={Number of papers},
    xlabel style={font=\footnotesize},
    ylabel style={font=\footnotesize},
    tick label style={font=\scriptsize},
    symbolic x coords={2024,2025,2026},
    xtick=data,
    ymin=0,
    ymax=28,
    ymajorgrids=true,
    grid style={opacity=0.3},
    nodes near coords,
    every node near coord/.append style={
        font=\scriptsize,
        /pgf/number format/fixed,
        yshift=2pt
    },
]
\addplot[
    fill={rgb,255:red,68;green,119;blue,170},
    draw=white,
    line width=0.5pt,
] coordinates {
    (2024,9)
    (2025,24)
    (2026,1)
};
\end{axis}
\end{tikzpicture}
\end{minipage}\hfill
\begin{minipage}[t]{0.46\textwidth}
  \centering
%

\begin{tikzpicture}
\begin{axis}[
    ybar,
    bar width=24pt,
    width=6cm,
    height=4.5cm,
    enlarge x limits=0.45,
    xlabel={Year},
    ylabel={Total citations},
    xlabel style={font=\footnotesize},
    ylabel style={font=\footnotesize},
    tick label style={font=\scriptsize},
    symbolic x coords={2024,2025,2026},
    xtick=data,
    ymin=0,
    ymax=280, 
    ymajorgrids=true,
    grid style={opacity=0.3},
    nodes near coords,
    every node near coord/.append style={
        font=\scriptsize,
        /pgf/number format/fixed,
        yshift=2pt
    },
]
\addplot[
    fill={rgb,255:red,238;green,119;blue,51},
    draw=white,
    line width=0.5pt,
] coordinates {
    (2024,236)
    (2025,54)
    (2026,0)
};
\end{axis}
\end{tikzpicture}
\end{minipage}
\Description{Left: bar chart showing 9 papers in 2024, 24 in 2025, and 1 in 2026.
Right: bar chart showing cumulative citation totals per year cohort as of mid-2025.}
\vspace{-8pt}
\caption{Temporal distribution of the reviewed corpus ($n=34$).
  \emph{Left:} number of included papers per publication year.
  \emph{Right:} Cumulative citation totals by publication year.}
\label{fig:temporal}
\end{figure}
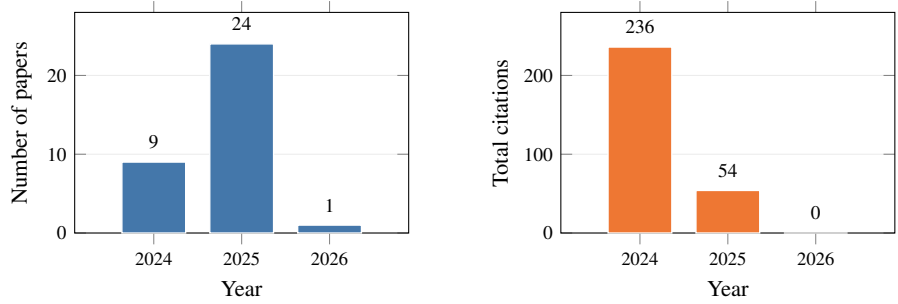

\begin{figure}[t]
  \centering
  \begin{minipage}[c]{0.33\textwidth}
    \centering
%
%
%
%
%
\resizebox{!}{3.5cm}{%
\begin{tikzpicture}[font=\footnotesize]

  \fill[blue!22]   (-1.10,  1.00) circle (1.90);
  \fill[red!18]    ( 1.10, -1.00) circle (1.55);

  \begin{scope}
    \clip (-1.10,  1.00) circle (1.90);
    \fill[violet!30] ( 1.10, -1.00) circle (1.55);
  \end{scope}

  \draw[blue!60!black, thick] (-1.10,  1.00) circle (1.90);
  \draw[red!55!black, thick]  ( 1.10, -1.00) circle (1.55);

  \node[blue!70!black, align=center] at (-1.10, 1.00)
    {{\large\bfseries 20}};

  \node[red!65!black, align=center] at (1.10, -1.00)
    {{\large\bfseries 13}};

  \node[violet!80!black, align=center] (intnode) at (0.15,-0.14)
    {{\large\bfseries 1}};

  \draw[gray!60, thin, dashed] (intnode.north east) -- (1.43,1.73);

  \node[align=center, font=\small, anchor=south] at (1.50,1.76)
    {\textit{Chelli et al.}\\[-0.5pt](2024)};

  \node[blue!70!black, align=center, anchor=south,
        font=\small\bfseries]
    at (-1.10,3.08)
    {Literature Retrieval};

  \node[red!60!black, align=center, anchor=north,
        font=\small\bfseries]
    at (1.10,-2.73)
    {Literature Screening};

\end{tikzpicture}}%
  \end{minipage}
  \hfill
  \begin{minipage}[c]{0.65\textwidth}
    \centering
%
%
\begin{tikzpicture}
\begin{axis}[
    xbar,
    bar width=7pt,
    width=6.2cm,
    height=4.8cm,
    xmin=0,
    xlabel={Number of papers},
    xlabel style={
        font=\footnotesize,
        yshift=4pt,
    },
    tick label style={font=\tiny},
    yticklabel style={
        anchor=east,
        align=right,
        font=\tiny
    },
    ytick={1,...,10},
    yticklabels={
        {Biomedical Text Mining and Ontologies},
        {Topic Modeling},
        {AI in Healthcare and Education},
        {Meta-analysis and Systematic Reviews},
        {Radiomics and ML in Medical Imaging},
        {Semantic Web and Ontologies},
        {Computational Drug Discovery Methods},
        {Natural Language Processing Techniques},
        {Machine Learning in Healthcare},
        {Advanced Graph Neural Networks}
    },
    y dir=reverse,
    enlarge y limits=0.05,
    xmajorgrids=true,
    grid style={opacity=0.3},
    nodes near coords,
    every node near coord/.append style={
        font=\tiny,
        /pgf/number format/fixed,
        anchor=west,
        xshift=2pt,
    },
]
\addplot[
    fill={rgb,255:red,68;green,119;blue,170},
    draw=white,
    line width=0.5pt,
] coordinates {
    (10,1)
    (7,2)
    (6,3)
    (6,4)
    (3,5)
    (3,6)
    (3,7)
    (3,8)
    (2,9)
    (2,10)
};
\end{axis}

\path[use as bounding box]
([yshift=16pt]current bounding box.south west)
rectangle
(current bounding box.north east);

\end{tikzpicture}
  \end{minipage}
    \vspace{-12pt}
    \caption{Overview of the reviewed literature. \emph{Left:} paper distribution across thematic categories. \emph{Right:} top 10 research topics by paper count.}

  \label{fig:overview}
\end{figure}
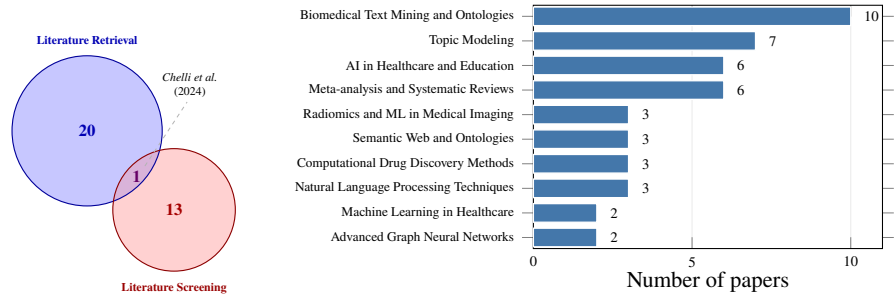

The experts organised the papers into two thematic categories (Figure~\ref{fig:overview}, left) by primary task: literature retrieval ($n=21$) and literature screening ($n=14$). One paper, \cite{Chelli2024}, appears in both categories.
Topic assignments from OpenAlex \citep{priem2022openalex} (Figure~\ref{fig:overview}, right) show that the corpus is concentrated in biomedical and clinical domains. The most frequent topic is ``Biomedical Text Mining and Ontologies'' ($10$ papers), followed by ``Topic Modeling'' ($7$), and ``AI in Healthcare and Education'' and ``Meta-analysis and Systematic Reviews'' ($6$ each). Overall, the distribution suggests that LLM-assisted literature workflows are being studied primarily in biomedical and evidence-synthesis settings.

\subsection{Presentation Approach}
The survey presents its two thematic areas in separate sections. Within each area, works are further organized into narrower groups defined by shared technical approaches or objectives.

Each group is discussed using the same three-part structure: (1) an \emph{Overview} summarizing the studies, their approaches, and the role of LLMs; (2) \emph{Models used}, covering which LLMs are employed, how they are accessed (e.g., API, local deployment) and whether they are fine-tuned; and (3) \emph{Evaluation trends}, covering how retrieval or screening quality is assessed, including the type of ground truth used, the metrics reported, and any indirect evidence from broader experiments.

It is worth to mention two important conventions used. First, \emph{Author-Constructed} ground truth refers to benchmarks, query sets, expert judgments, or synthetic data created specifically for the reviewed work; datasets introduced in earlier separate publications are treated as \emph{Existing Benchmarks}, even if the author lists overlap. Second, only evaluations of LLM-assisted retrieval or screening components are reported: results for unrelated pipeline stages and results where LLMs appear only as baselines are excluded.

\subsection{Limitations }
\label{sec:corpus:limits}

Several potential sources of non-representativeness should be acknowledged.
First, the search was conducted through a single scholarly search engine (BIP!~Finder), which, while it has a broad coverage, may not index all venues uniformly.
Second, the restriction to English-language publications excludes relevant work published in other languages.
Third, the rapidly evolving nature of the field means that the corpus represents a snapshot: work published after the data collection date (see Section~\ref{sec:corpus:search}) is not included.
Moreover, the survey is restricted to peer-reviewed publications, hence, preprints that had not yet undergone peer review at the time of screening (even if they are subsequently accepted and published) are not included. Finally, the inclusion criteria require explicit LLM usage as a primary method for retrieval or screening, meaning that important adjacent work is out of scope.

Despite these limitations, the corpus provides a reasonably comprehensive view of the state of the field, capturing the primary methodological strands, the recent contributions, and the principal application domains in which LLM-based literature retrieval, recommendation, and screening have been deployed.

\section{LLMs for Literature Retrieval}
\label{sec:retrieval_recommendation}

Identifying relevant scientific literature underpins most research activities, yet established methods are increasingly strained by the growth described in Section~\ref{sec:introduction}. Effective querying requires anticipating shifting terminology, assessing relatedness requires semantic and conceptual understanding beyond surface overlap, and judging novelty or evidential relevance at scale exceeds what keyword matching or citation counts can reliably capture. Generative LLMs are therefore a natural fit for these challenges. The papers reviewed next fall into two broad groups: those where retrieval or recommendation is the primary contribution, and those where retrieval supports a downstream task such as automated peer review, clinical decision support, or survey writing.

\subsection{Core Literature Search \& Recommendation}
\label{sec:core-search}

\noindent\textbf{Overview.} This is the largest and most technically diverse area identified in the survey’s literature-retrieval domain. It includes systems that retrieve scientific documents relevant either to an explicit natural-language or keyword-based query or to a stored profile of a user’s past interactions. We identify five distinct research strands within it.

The first strand concerns semantic search. \emph{SemRank} \citep{Zhang2025b} addresses a limitation of dense retrievers: by encoding queries and passages holistically, they can miss the fine-grained scientific concepts that determine relevance. It uses an LLM to extract multi-granular concepts at indexing time and again at query time to identify the concepts the user is actually seeking, matching against a concept-based index rather than raw embeddings alone. \emph{Pathfinder} \citep{Iyer2024} combines retrieval-augmented generation with time- and citation-based weighting over astronomy papers, allowing natural-language querying and embedding-space exploration; an LLM then performs consensus evaluation over the retrieved literature. \citet{Zheng2025} similarly treats relevance as a constrained optimization problem, using an LLM-conditioned encoder to embed queries and documents, then re-weighting their cosine similarity with constraints that penalize redundancy and reward semantic and logical consistency.
\emph{BioMedSearch}~\citep{Liu2025b} uses an LLM to decompose a biomedical question into sub-queries and keywords, builds a task graph, and fans out simultaneously across literature databases, protein databases, and general web search, applying a keyword-coverage threshold and embedding-similarity ranking to narrow the combined multi-source results before the LLM reasons over them. \citet{deArajoPessoa2025} builds a RAG pipeline for institutional publication analysis, where an LLM converts natural-language requests into full-text search queries and combines vector search with collaboration-network analysis. \emph{RefAI} \citep{Li2024a} similarly uses an LLM to turn user queries into PubMed keywords, then ranks results using citation and journal-impact signals alongside semantic relevance. Finally, \citet{YaoDing2025} reports using an LLM to embed patent titles, abstracts, and technical descriptions for retrieval.

A second strand foregrounds the agentic dimension of search, using LLMs not just to encode meaning but to plan, route, and refine retrieval itself. 
\emph{TourSynbio-Search} \citep{Liu2024} use a classification layer with chain-of-thought (CoT) and few-shot prompting to decide, per query, whether to route to a paper-search agent or a protein-structure-search agent. \emph{LitChat} \citep{Huang2025} splits the problem across two cooperating agents: one that turns a conversational request into a Boolean query executable against Web of Science, Scopus, or Semantic Scholar (using in-context examples drawn from real systematic-review search strategies), and a second that mines the retrieved corpus, organized as a bibliographic knowledge graph, to answer follow-up questions. Finally, \emph{LITERAS}~\citep{Gorenshtein2025} introduces a four-agent system: a Keyword Planner Agent generates PubMed-ready search strings, a Search Agent queries the PubMed API and deduplicates results, a Validation Agent filters and ranks the retrieved articles against relevance, recency, methodology, and quality criteria, and a Critic Agent loops back to refine the search if too few articles clear the threshold.

A third strand grounds retrieval in structured knowledge rather than free-text similarity. \emph{GPTscholar}~\citep{Pires2024} has the LLM generate a SPARQL query against the DBLP bibliographic knowledge graph before answering, explicitly adopting a ``knowledge-aware inference'' strategy to suppress hallucinated titles, authors, and DOIs. 
\citet{Liu2025a} takes a structurally similar approach in the opposite direction — it first builds the knowledge graph itself, via a novel ontology and an LLM-augmented multi-relation extraction model, then lets GPT-4o generate Cypher queries against the resulting Neo4j graph to answer content-level literature questions, arguing this substrate supports deeper exploration than keyword search or even a bare LLM.

The fourth strand focuses not on retrieving relevant papers, but on the accuracy of citations that an LLM generates from its parametric memory without grounding them in a database query 
\citet{Chelli2024} conducted a controlled study in which ChatGPT and Bard/Gemini were asked to reproduce the reference lists of systematic reviews on rotator-cuff pathology. 

A fifth strand departs from query-driven search entirely, framing literature discovery as profile-driven literature recommendation. \emph{MCAP}~\citep{Zhang2025a} represents users and papers as nodes in heterogeneous interaction graphs, and argues that the high-order message-passing propagation used by prior graph neural network (GNN) recommenders is sub-optimal, since long-distance propagation blurs useful signal with noise. Its fix (low-pass propagation with matrix completion) builds direct user–user and item–item relation graphs (from shared authorship/venue and content similarity) instead of relying on noisy multi-hop connections. LLMs (GPT-3.5-Turbo, GLM-4) play only a narrow, auxiliary role here: refining which papers count as ``similar'' for the item–item graph; the GNN recommender, trained from scratch, is the paper's real contribution.

\noindent\textbf{Models used.}
In this area, many systems rely on general-purpose commercial models used off the shelf, most commonly via API, though access method is frequently left unstated. \citet{Pires2024} uses GPT-3.5-turbo variants, \citet{Li2024a} uses GPT-4-turbo, and \citet{Huang2025} and \citet{Liu2025a} use GPT-4o, all with API access explicitly confirmed. \citet{Zhang2025b} uses GPT-4.1-mini for concept extraction, although, as in \citet{YaoDing2025}, the paper does not specify the access method. \citet{Liu2025b} compares five models side by side: GPT-4.1, DeepSeek-R1, Gemini-2.5, Llama-4, and Qwen3. \citet{deArajoPessoa2025}, by contrast, uses the open-weight Llama-3.3-70B model through the UniGPT API service.  
\citet{Iyer2024} uses GPT-4o-mini for its post-retrieval consensus-checking module. \citet{Chelli2024} is the only Web UI-based study, testing GPT-3.5, GPT-4, and Bard/Gemini directly through their consumer chat interfaces rather than via API. 
Across these studies, the models are used without fine-tuning; \citet{Zhang2025a} follows the same pattern for its LLM components (GPT-3.5-Turbo, GLM-4), though these are used only to generate semantic side-information for a graph-based recommender rather than to perform retrieval directly. 
Domain-specific LLM fine-tuning appears in only one paper: TourSynbio-Search fine-tunes InternLM2-7B into TourSynbio-7B on curated protein literature and deploys it locally. 
Finally, \citet{Zheng2025} is the outlier that never names a concrete model. Columns 2-3 of Table~\ref{tab:core-search} summarise the previous discussion.

\begin{table}[t]
\caption{Models, Access Modes, Retrieval-related Evaluation Ground Truth, and Metrics for Core Literature Search \& Recommendation Approaches. N/R = Not reported}
\label{tab:core-search}
\centering
{\scriptsize
\setlength{\tabcolsep}{3pt}
\begin{tabular}{p{2.2cm}p{3.2cm}lp{1.2cm}p{3.5cm}}
\hline\noalign{\smallskip}
\textbf{Paper} & \textbf{LLM Models} & \textbf{Access} & \textbf{Gr. Truth} & \textbf{Evaluation} \\
\noalign{\smallskip}\svhline\noalign{\smallskip}
\citet{Iyer2024} [Pathfinder] & GPT-4o-mini & N/R & - 
& No LLM-assisted retr. evaluation \\
\hline
\citet{Zhang2025b} [SemRank] & GPT-4.1-mini & N/R & Existing benchmark& Recall@K, Time, Cost \\
\hline
\citet{Zheng2025} & N/R & N/R & Existing benchmark& Precision@K, Recall@K, NDCG@K, MRR \\
\hline
\citet{Liu2025b} [BioMedSearch] & GPT-4.1, Llama-4, Gemini-2.5, DeepSeek-R1, Qwen3 & API & - & No retr. evaluation \\
\hline
\citet{deArajoPessoa2025} & Llama-3.3-70B & API & - & No retr. evaluation \\
\hline
\citet{Li2024a} [RefAI] & GPT-4-Turbo & API & Author-constructed & Expert satisfaction \\
\hline
\citet{YaoDing2025} & GPT-3.5-turbo, GPT-4 (as baseline comparison) & N/R & N/R & Accuracy, Recall \\
\hline
\citet{Liu2024} [TourSynbio-Search]& TourSynbio-7B (fine-tuned from InternLM2-7B on curated ProteinLMDataset) & Local & - & No retr. evaluation \\
\hline
\citet{Huang2025} [LitChat] & GPT-4o & API & - & No retr. evaluation \\
\hline
\citet{Gorenshtein2025} [LITERAS] & GPT-4o-mini & API & Author-constructed & Hallucination rate, Time (limited)\\
\hline
\citet{Pires2024} [GPTscholar] & GPT-3.5-turbo & API & Author-constructed & Correctness counts, Hallucination rate \\
\hline
\citet{Liu2025a} &
GPT-4o & API & Author-constructed & Expert satisfaction, Time (subjective) \\
\hline
\citet{Chelli2024} & GPT-3.5/4, Gemini (Bard) & Web UI& Existing syst. reviews & Precision, Recall, F1, Hallucination rate (based on simul. prompt)\\
\hline
\citet{Zhang2025a} [MCAP] & GPT-3.5-Turbo; GLM-4 & N/R & Existing benchmark & Recall@K, NDCG@K, Hit Ratio@K \\
\hline
\end{tabular}
}
\end{table}

\noindent\textbf{Evaluation trends.}
Many papers in this line of work evaluate literature retrieval with classic information-retrieval metrics (e.g., Recall, Precision, nDCG), reflecting that finding relevant documents is these papers' primary focus. These metrics are calculated against existing open benchmarks \citep{Zhang2025b,Zhang2025a,Zheng2025} or existing systematic reviews~\citep{Chelli2024}.\footnote{\citet{Chelli2024} does not involve genuine retrieval: ChatGPT and Bard generate citations from parametric memory without querying any database or corpus. Its Precision/Recall/F1 therefore measure overlap with a real reference list, not retrieval quality.} 
\citet{Zhang2025a} specifically evaluates against established recommender-systems benchmarks, using standard top-$k$ ranking metrics (Recall@5, NDCG@5, Hit Ratio@5). \citet{Gorenshtein2025} and \citet{Pires2024} rely on manual quality and hallucination-frequency ratings on author-constructed prompt sets rather than ranking metrics. \citet{Li2024a} similarly evaluates against an author-constructed set of queries, using expert Likert-scale ratings of relevance and quality. \citet{Iyer2024} reports pipeline-level recall, nDCG, and MRR on an author-constructed ground truth, but the LLM component itself is not directly evaluated: its contribution, consensus evaluation of the retrieved set, is peripheral to these metrics and is not separately assessed.
Four papers \citep{Liu2025b,Liu2024,Huang2025,deArajoPessoa2025} report no formal quantitative evaluation of the core retrieval task at all, even though most of them do include other, non-retrieval-related experiments.
Finally, cost and processing-time reporting is rare and mostly informal: \citet{Zhang2025b} stands out with a dedicated efficiency analysis (retriever/LLM calls per query, average LLM output length, running time, and actual dollar cost), while \citet{Gorenshtein2025} reports measured average processing time, and \citet{deArajoPessoa2025} and \citet{Liu2025a} offer only qualitative or subjective time/latency observations.
Columns 4–5 of Table~\ref{tab:core-search} summarise the previous discussion.

\subsection{RAG-Style Literature Retrieval for Downstream Tasks}
\label{sec:rag}

\noindent\textbf{Overview.}
All papers in this area adopt some form of RAG (Retrieval-Augmented Generation) combining literature retrieval with a downstream generation or reasoning stage conditioned on the retrieval material. However, only a subset explicitly use the term: \citet{Xu2025, Fu2025}, 
and \citet{Bao2025} self-describe as RAG, while \citet{Peasley2025} and \citet{Zhu2025} use the term only in passing, and \citet{Zheng2026} and \citet{Balaskas2024} never use it at all despite following the identical retrieve-then-condition pattern. Consistent with this survey's scope, we examine only the literature-retrieval component of these pipelines here; the quality of the downstream generation itself (e.g., review text, treatment plan, survey prose) is not covered.

We identify two strands of works distinguished by the type of information retrieved. The related- and concurrent-work retrieval strand retrieves prior papers to contextualize a target paper's novelty or limitations. 
\emph{DeepReview}~\citep{Zhu2025} builds its retrieval stage on OpenScholar: Qwen-2.5-72B-Instruct generates key research questions about the target paper, Qwen-2.5-3B-Instruct converts these into search keywords used to pull roughly $60$ candidate papers via the Semantic Scholar API, after which OpenScholar's own (non-generative, cross-encoder) reranker narrows this to the top $10$ most relevant papers and its QA model synthesizes the resulting novelty-analysis report. \emph{LIMITGEN}~\citep{Xu2025} augments LLMs with RAG over the same Semantic Scholar API to ground generated limitations in prior findings. Its premise is that identifying weaknesses in a paper (e.g., missing baselines, limited datasets) requires literature awareness beyond that of a static LLM. However, the authors position their contribution primarily as a benchmark for assessing LLMs in this setting, rather than an optimized retrieval approach. \emph{PaperEval}~\citep{Zheng2026} takes a more self-contained approach: ChatGPT generates representative topic keyphrases for a target paper, which are encoded via a CLIP text encoder. Cosine similarity against a corpus (restricted to concurrent, temporally-proximate publications) then retrieves a domain-aware reference set that a latent-reasoning module uses to assess the paper's novelty and contribution. 

By contrast, the domain-literature evidence retrieval strand retrieves general supporting evidence to inform a downstream generation or decision task.  
\emph{LITURAt}~\citep{Peasley2025} is fundamentally a ``plan-and-solve'' scientific-data-analysis agent, built around Mixtral 8×7B, that dynamically constructs Entrez API queries to retrieve relevant PubMed abstracts alongside local datasets so its statistical summaries are contextualized by current literature. \citet{Fu2025} goes further, building an explicit ``clinical data–literature knowledge–model decision-making'' collaborative framework. The system combines RAG-retrieved oncology literature with structured patient data through a staged, fine-tuned 70B medical LLM. The system in \citep{Balaskas2024} integrates a similar approach directly into an electronic health record interface: WizardVicuna-13B first generates a natural-language summary from a patient's structured data; this is then embedded and used to query the literature. Dense vector retrieval returns $100$ candidate documents, which a cross-encoder reranks to the top $10$. A relevance threshold then determines whether any results are shown to clinicians. The retrieval criteria explicitly prioritise patient applicability, treatment alignment, and source recency and credibility. \emph{SurveyGen/QUAL-SG}~\citep{Bao2025} identifies literature retrieval as one of two key bottlenecks in existing RAG-based survey pipelines (the other being evaluation): standard retrieval selects papers by textual similarity alone and risks including low-impact or marginal work, so QUAL-SG expands the candidate pool via citation-graph analysis and re-ranks by three signals, including an LLM-judged (Claude-3.7-Sonnet) relevance score, before handing the retrieved literature to the generation stage.

\noindent\textbf{Models used.} 
The works in this area draw on a genuinely diverse set of models. Regarding LLM access, \citet{Xu2025} uses an API,\footnote{Based on the information in the author's code repository.} while \citet{Peasley2025} and \citet{Balaskas2024} instead use locally deployed open-weight models: Mixtral 8×7B and WizardVicuna-13B, respectively. The remaining studies—\citep{Zhu2025}, \citep{Zheng2026}, \citep{Fu2025}, and \citep{Bao2025}-do not explicitly describe how their models are accessed or deployed. 
Regarding the models, \citet{Zhu2025} uses three generative LLMs in its retrieval pipeline: Qwen-2.5-72B-Instruct for research questions generation, Qwen-2.5-3B-Instruct for search keywords generation, and Llama-3.1\_OpenScholar-8B (fine-tuned from Llama3-8B) for novelty-report synthesis. The pipeline also uses OpenScholar’s cross-encoder, a non-generative reranker fine-tuned from BAAI/bge-reranker-large, to reduce the retrieved candidate pool to the top 10.\footnote{The work additionally uses other LLMs for tasks that do not touch literature retrieval, including DeepReviewer-14B (paper's full-parameter fine-tune of Phi-4 14B).} \citep{Xu2025} tests four frontier LLMs head-to-head, zero-shot: GPT-4o, GPT-4o-mini, Llama-3.3-70B, and Qwen2.5-72B. \citet{Zheng2026} names only ChatGPT (version unspecified) for keyphrase generation in its retrieval module\footnote{GPT-4o and GPT-4o-mini appearing elsewhere in that paper belong to separate baseline methods, not to PaperEval's own pipeline} while the backbone of PaperEval's own reasoning module is never named. \citet{Fu2025}'s 70B ``Zhuomuniao'' model has an undisclosed base architecture, fine-tuned in three stages (LoRA domain adaptation, contrastive subtype discrimination, RLHF-based contraindication reduction). \citet{Bao2025}'s retrieval pipeline is anchored by Claude-3.7-Sonnet, which performs the LLM-judged topical-relevance scoring inside QUAL-SG itself and is also the paper's best-performing and most extensively used model overall, spanning all three evaluation tasks. Columns 2-3 of Table~\ref{tab:rag} summarise the previous discussion.

\begin{table}[t]
\caption{Models, Access Modes, Retrieval-related Evaluation Ground Truth, and Metrics for RAG-Style Lit. Retrieval for Downstream Tasks Approaches. N/R = Not reported}
\label{tab:rag}
\centering
{\scriptsize
\setlength{\tabcolsep}{3pt}
\begin{tabular}{p{1.9cm}p{3.4cm}lp{1.3cm}p{3.4cm}}
\hline\noalign{\smallskip}
\textbf{Paper} & \textbf{LLM Models} & \textbf{Access} & \textbf{Gr. Truth} & \textbf{Evaluation} \\
\noalign{\smallskip}\svhline\noalign{\smallskip}
\citet{Zhu2025} [DeepReview]& Qwen-2.5-3B-Instruct; Qwen-2.5-72B-Instruct; Llama-3.1\_Open\-Scholar-8B (fine-tuned from Llama3-8B) & N/R & - & No retr. evaluation; Cost reporting (examining tradeoffs) \\
\hline
\citet{Xu2025} [LIMITGEN]& GPT-4o; GPT-4o-mini; Lla-Ma-3.3-70B; Qwen2.5-72B & API &  - & No retr. evaluation (indirect signal from RAG-context ablation) \\
\hline
\citet{Zheng2026} [PaperEval] & GPT & N/R & - & No retr. evaluation (indirect signal from ranking ablations)\\
\hline
\citet{Peasley2025} [LITURAt] & Mixtral-8x7B & Local & - & No retr. evaluation (indirect signal from the combined report stability experiment)\\
\hline
\citet{Fu2025} & Zhuomuniao-70B (undisclosed base architecture) & N/R & - & Recall/Precision Proxies (e.g., percentage of high-impact literature); Time (end-to-end proc. times)\\
\hline
\citet{Balaskas2024} & WizardVicuna-13B (fine-tuned from LlaMa-13B) & Local & - & No retr. evaluation \\
\hline
\citet{Bao2025} [QUAL-SG] & Claude-3.7-Sonnet & N/R & Author-constructed & Precision, Recall, F1 \\
\hline
\end{tabular}
}
\end{table}

\noindent\textbf{Evaluation trends.} 
The dominant pattern is that literature-retrieval quality is rarely evaluated against a dedicated ground truth: six of the seven papers report no retrieval-specific ground truth, relying instead on indirect signals or ground truth scoped to the downstream task. This likely reflects where these papers place their contribution: retrieval is a means to an end (e.g., grounding a review, a treatment plan, a patient summary) so evaluation effort naturally concentrates on the quality of that downstream output rather than on the retrieval step that feeds it. \citet{Zhu2025} reports no retrieval-relevant signal of any kind; even its LLM-as-judge review-quality scores are holistic ratings of the final review text, with no rubric item or ablation attributing any part of the judgment to retrieval specifically. \citet{Xu2025} and \citet{Zheng2026} each provide an indirect signal via ranking-position ablations: \citet{Xu2025} varies which subset of already-retrieved papers is used downstream (Top 5 vs. Top 3 vs. Last 5 of 18), and \citet{Zheng2026} both removes its retrieval module entirely (``w/o Ret.'') and sweeps the number of retrieved papers $k$; however, neither compares retrieved content against a labeled relevant-document set, so both remain indirect signals rather than genuine retrieval evaluations. \citet{Peasley2025} provides a similar indirect signal. The authors limit their only ground-truth accuracy evaluation to the local-data-analysis module, noting that they could not establish ground truth for the literature review. Hence,  insights come from an internal/external consistency experiment (Cohen's $\kappa$, self-judged by Mixtral) on the combined report. 
\citet{Fu2025} reports what we term recall/precision proxies. Compared with a no-RAG baseline, its RAG-enabled system achieves a 42.3\% increase in citations to recent literature, serving as a proxy for recall or coverage, and cites a higher proportion of high-impact literature, serving as a proxy for precision or quality. Both measures are computed from the system’s own outputs rather than evaluated against an external ground truth of relevant documents. 
\citet{Balaskas2024} is a clean no-evaluation case, reinforced by the authors' own admission that retrieval-quality assessment is left to future work.
\citet{Bao2025} is the only paper with genuine, validated retrieval ground truth: it builds a dedicated author-constructed dataset ($4,200$+ human-written surveys with real citations) and reports citation Precision/Recall/F1 against human-selected references. A four-component ablation further isolates each ranking signal's individual contribution: removing the LLM-judged relevance component (Claude-3.7-Sonnet) drops citation F1 by $4.44$ points (from $16.73$ to $12.29$). QUAL-SG's retrieval quality is also validated against dedicated re-ranking baselines (RankGPT, UPR).
Cost/time reporting is rare but not entirely absent: no paper reports a retrieval-specific cost or timing figure, but \citet{Zhu2025} reports a total token-generation/inference-cost tradeoff and \citet{Fu2025} reports end-to-end system processing times. Columns 4-5 of Table~\ref{tab:rag} summarise the previous discussion.

\section{LLMs for Literature Screening}
\label{sec:screening}

Literature screening, the classification of candidate publications against predefined eligibility criteria, is widely seen as the most labor-intensive stage of systematic reviews and related evidence-synthesis workflows. The LLM-based papers reviewed here split into two broad lines of work: some treat screening as a single model call per paper, with the main technical variation lying in prompt design, while others use multi-component systems intended to move beyond one-shot classification toward more practical screening tools. The following sections present these two areas in turn.

\subsection{Single-Pass LLM Screening}
\label{sec:screen1}

\noindent\textbf{Overview.} This group treats the screening decision as the output of a single, independent model call per candidate paper: one model, one prompt, one classification or, in \citet{Chelli2024}'s case, one generative response. Several papers in the group involve more than one LLM, but the models are compared, not combined. There is no chaining, no retrieval infrastructure, and no inter-model arbitration deciding the final answer. This architectural simplicity means these papers function largely as benchmarking studies, establishing how well an off-the-shelf LLM call performs at screening.

The real technical variation across this group lies entirely in how the prompt and evaluation protocol are constructed, and the spread is considerable. At the simplest end, \citet{Chelli2024} and \citet{Delgado-Chaves2025} use fixed, static criteria-statement prompts: \citet{Chelli2024} tests only whether specifying a minimum number of papers to retrieve changes the model's output, while \citet{Delgado-Chaves2025} holds the prompting technique constant across $18$ models and instead demonstrates that the formulation of the inclusion/exclusion criteria themselves is a major performance driver, independent of model choice. \citet{Ronquillo2024} moves one step further by treating the prompt itself as the experimental variable: the same three models are each run through four escalating prompt-optimization variants, directly measuring how much structure added to the prompt (rather than which model receives it) moves accuracy.

\citet{Li2024b} pushes prompt engineering further still, using few-shot examples paired with explicit reasoning explanations and running a systematic ablation across two axes: shot count (0-shot through 10-shot) and the source of the reasoning explanations attached to each example, comparing human-expert-written versus LLM-generated reasoning as a variable in its own right. 
\citet{Tao2025} develops this same idea into the most elaborate prompt-engineering process in the group: although the deployed screening call is still a single model (GPT-4o, Kimi, or DeepSeek, tested separately), the prompt behind that call is built offline through an iterative, three-phase process: a small seed set of PICOS-derived few-shot examples, progressive expansion with wording refinement, and finally folding misclassified cases directly back in as new few-shot examples, an active-learning-style loop that shapes the prompt without ever changing the underlying single-call architecture. \citet{Ruan2025} takes a further, distinctive detour from hand-designed prompting altogether, using meta-prompting (asking the LLMs themselves to draft candidate screening prompts), iteratively refining the machine-suggested prompts before applying them uniformly across all four models being compared. \citet{Oami2025} use a more conventional fixed role/PICOS-structured prompt but add a methodological check: the identical prompt is run three times to test reproducibility, and a post hoc ensemble rule (include if flagged by either of two models) is explored as a secondary analysis. 

\noindent\textbf{Models used.} Most papers in this group use commercial chat-model families: OpenAI's GPT-3.5/GPT-4/GPT-4o, Anthropic's Claude (2 and 3.5), and Google's Gemini/Bard. \citet{Delgado-Chaves2025} stands out as the exception: alongside GPT-3.5-turbo, GPT-4o, and GPT-4o-mini, they run a genuinely broad sweep of $15$ open-source models spanning Google's Gemma/Gemma2 family, Meta's Llama3/3.1/3.2 line, Mistral AI's Mistral/Mistral-Nemo/Mixtral models, and Qwen2.5-7B, $18$ models in total, making it the widest single-paper model comparison in the whole set. Two of these open-source entries, Reflection-70B and Athene-70B, are themselves third-party fine-tunes of Llama (by Reflection AI and Nexusflow, respectively); the authors though do not fine-tune any model, and this holds across the whole group: all papers use off-the-shelf models, with ``tuning'' limited to prompt-level techniques rather than gradient-based adaptation of the underlying model. Access is mostly API-based, except for \citet{Delgado-Chaves2025}, which runs open-source models locally via Ollama, and \citet{Chelli2024}, which uses the web chat UI instead of an API.  Columns 2-3 of Table~\ref{tab:screening_4SR_summary} summarise the previous discussion.

\begin{table}[t]
\caption{Models, Access Modes, Screening Evaluation Ground Truth, and Metrics for Single-Pass LLM Screening Approaches. N/R = Not reported}
\label{tab:screening_4SR_summary}
\centering
{\scriptsize
\setlength{\tabcolsep}{3pt}
\begin{tabular}{p{1.1cm}p{4.2cm}lp{1.2cm}p{3.7cm}}
\hline\noalign{\smallskip}
\textbf{Paper} & \textbf{LLM Models} & \textbf{Access} & \textbf{Gr. Truth} & \textbf{Evaluation} \\
\noalign{\smallskip}\svhline\noalign{\smallskip}

\citet{Chelli2024} & 
GPT-3.5/4; Gemini (Bard) & Web UI & Existing syst. reviews & Precision, Recall, F1, Hallucination rate (based on simul. prompt) \\
\hline
\multirow{2}{1.1cm}{\citet{Delgado-Chaves2025}} &
GPT-3.5-turbo, GPT-4o, GPT-4o-mini  &
API &
\multirow{2}{1.2cm}{Existing syst. reviews} &
\multirow{2}{3.7cm}{Precision, Recall, Specificity, F1, MCC, PABAK} \\
& Gemma-7B; Gemma2-9B/27B; Llama3-8B/-8B-instruct-fp16; Llama3.1-8B/70B; Llama3.2-3B; Reflection-70B (Llama-based); Athene-70B (Llama-based); Mistral-v0.2/v0.3-7B; Mistral-Nemo-12B; Mixtral-8x22B; Qwen2.5-7B &
Local &
&
\\
\hline
\citet{Li2024b} &
GPT-3.5; GPT-4; Claude-2 &
API &
Existing benchmark &
Recall, Specificity, Precision, F1, Consistency (of multiple runs) \\
\hline
\citet{Ronquillo2024} &
GPT-3.5 Turbo; GPT-4; Claude-2 &
API &
Author-constructed &
Accuracy (chi-square comparisons) \\
\hline
\citet{Ruan2025} &
\begin{tabular}[t]{@{}l@{}}
GPT-4; Claude-3.5;
Gemini-1.5; \\DeepSeek-V3
\end{tabular} &
N/R &
Author-constructed & Recall- and Specificity-based metrics (FNF, FPF, PLR, Youden's Index, NNS), Risk difference, Risk ratio, Redundancy number, Screening time\\
\hline
\citet{Oami2025} &
GPT-4o; Claude-3.5-Sonnet; Gemini-1.5-Pro; Llama-3.3-70B &
API &
Existing syst. reviews &
Recall, Specificity, Scree\-ning time, Cost, Consistency (of multiple runs) \\
\hline
\citet{Tao2025} &
GPT-4o; Kimi (Moonshot-v1-128k); DeepSeek-Chat 2.5 &
API &
Author-constructed &
Accuracy, Precision, Recall, Exclusion-reason concordance rate, Time \\
\hline
\end{tabular}
}
\end{table}

\noindent\textbf{Evaluation trends.}
Ground truth sources in this group include existing systematic reviews used as a known answer key \citep{Chelli2024,Delgado-Chaves2025,Oami2025}, author-constructed ground truth built specifically for the study \citep{Ronquillo2024,Ruan2025,Tao2025} and one pre-existing, independently curated benchmark dataset \citet{Li2024b}. 
Regarding the used metrics, this group leans heavily on the classic diagnostic-accuracy toolkit: precision, recall/sensitivity, specificity, and F1 appear in nearly every paper, supplemented in \citep{Delgado-Chaves2025} by agreement statistics that correct for class imbalance (MCC and PABAK). Beyond this common core, several papers introduce task-specific or efficiency-oriented metrics that reflect their particular angle: \citet{Chelli2024}'s hallucination rate (since its task is generative rather than classificatory); \citet{Li2024b}'s and \citet{Oami2025}'s consistency metrics, both measuring agreement between repeated runs (computed by varying the few-shot examples between runs in the former and by simply rerunning the identical prompt in the latter); \citet{Ruan2025}'s false-negative and false-positive fractions (mathematically equivalent to 1 minus sensitivity and 1 minus specificity, respectively), reported alongside a set of diagnostic-performance measures borrowed from the clinical diagnostic-test tradition (PLR, Youden’s Index, NNS, risk difference, risk ratio) and a paper-specific redundancy number tracking the manual re-review burden left after screening (computed for a non-LLM comparator, RobotSearch, as well as for the four LLMs); and \citet{Oami2025}'s and \citet{Tao2025} shared emphasis on cost and processing-time efficiency alongside accuracy. Columns 4-5 of Table~\ref{tab:screening_4SR_summary} summarise the previous discussion.

\subsection{Engineered Multi-Component Pipelines}
\label{sec:screen2}

\noindent\textbf{Overview.} This group treats the screening decision itself as the output of a designed system with multiple interacting components: pre-filtering stages, retrieval infrastructure, inter-model arbitration, or chained/tiered prompts. Critically, in every paper here the multi-component architecture is built specifically to produce the include/exclude decision, not some other downstream task; where a paper also performs data extraction or synthesis afterward, that later stage runs as a separate process on the already-screened subset and is excluded from this discussion. The shared ambition across this group is to move beyond a one-shot classification benchmark toward something closer to a production-usable screening tool, generally by reproducing the checks and balances a human review team would apply (e.g., a second reviewer, a tie-breaking adjudicator, a retrieval step) rather than relying on a single model's raw judgment.

The lightest form of this engineering is a pre-filter-then-classify cascade. \citet{Ji2024}'s \emph{ModelDB} pipeline narrows a large corpus using SPECTER2 embeddings and k-nearest-neighbours distance before GPT-3.5/GPT-4 confirm inclusion criteria on the reduced set, with temperature and chain-of-thought (CoT) reasoning further tuned on top of the confirmation step; the two stages work together purely to produce the relevance decision, with any subsequent metadata extraction handled as a separate step on the papers that already passed screening.
A step up in complexity is explicit inter-model arbitration, where multiple models screen independently but a designated component resolves their disagreements rather than leaving that to a human or a simple majority vote. \citet{Dai2025} has two models (ChatGPT-4o, Claude-3.5 Sonnet) screen in parallel, with a third model, Gemini-1.5 Pro, explicitly arbitrating any disagreements between them, a genuine consensus mechanism, further refined via a post hoc, error-driven CoT prompt revision cycle targeting identified false-negative patterns. \citet{Pan2025} builds a heavier version of the same idea: a tripartite framework (Doubao-1.5-pro-32k, DeepSeek-v3, DeepSeek-R1-Distill-Qwen-7B) in which the third model is purpose-built as an adjudicator, resolving conflicting inclusion/exclusion decisions between the other two and producing supporting reasoning for its ruling.

A second architectural pattern is retrieval-augmented screening, where the system doesn't just prompt a static context but actively retrieves information to answer the screening question. \citet{Trad2025} pairs a prompt-engineered, role-based title/abstract stage (with a conservative yes/no/unsure decision rule that retains uncertain articles) with a full-text screening stage built on retrieval-augmented generation (full-text PDFs are indexed into a vector store, and GPT-4 retrieves relevant passages to answer eligibility questions rather than relying on a fixed prompt alone). This is architecturally distinct from the arbitration papers above: rather than combining multiple models' judgments, it's a single model whose screening decision is scaffolded by a dedicated retrieval component. 
\citet{Lin2025}'s decision-tree pipeline for precision-oncology curation follows a related but distinct chaining logic: prompts are organized into sequential tiers, each tier's output feeding the next, with each stage referencing external regulatory databases (ARTG, PBS, FDA) to assign a tiered evidence classification.

The most elaborate systems in this group are explicitly agentic and iterative, with the multi-agent mechanism dedicated entirely to the screening questions. \citet{Hu2025}'s \emph{GREP-Agent} integrates four LLMs with a dedicated fine-tuning phase, in which human reviewers provide feedback on a small labelled subset to iteratively refine the screening prompts before an operational phase applies workload-reduction thresholds to decide when human review of a screening decision is still needed. The entire architecture, from fine-tuning to operation, is scoped to the three inclusion/exclusion questions the system answers. \citet{Mourtzinis2025}  similarly requires consensus across four LLMs on each of six screening criteria, with human experts stepping in specifically to break ties.

\noindent\textbf{Models used.}
The works in this group combine commercial and open-source models within each system rather than benchmarking many models against one another. GPT-4/GPT-4o/GPT-4o-mini and Claude 3.5 Sonnet appear in the arbitration- and retrieval-based systems of \citep{Dai2025,Trad2025,Hu2025}. \citet{Lin2025} uses a single fully local open-source model, Mistral-Nemo-12B, while \citet{Pan2025,Mourtzinis2025} favor newer open-weight families. \citet{Hu2025}’s GREP-Agent mixes GPT-4o/GPT-4o-mini with Llama-3.1 and Phi-4. Similarly to the approaches presented in the previous section, none of these papers fine-tunes model weights. 
Access is confirmed as API-based for \citep{Ji2024}, \citep{Trad2025}, and \citep{Pan2025}; \citet{Lin2025} instead uses a local deployment, while \citet{Dai2025} uses a Web UI. Access is not explicitly reported for \citep{Hu2025} or \citep{Mourtzinis2025}, despite their automated, multi-model pipelines making API access plausible. Columns 2-3 of Table~\ref{tab:screening_asInfra_summary} summarise the previous discussion.

\begin{table}[t]
\caption{Models, Access Modes, Screening Evaluation Ground Truth, and Metrics for Engineered Multi-Component Pipelines. N/R = Not reported}
\label{tab:screening_asInfra_summary}
\centering
{\scriptsize
\setlength{\tabcolsep}{3pt}
\begin{tabular}{p{1.6cm}p{3.7cm}lp{1.4cm}p{3.2cm}}
\hline\noalign{\smallskip}
\textbf{Paper} & \textbf{LLM Models} & \textbf{Access} & \textbf{Gr. Truth} & \textbf{Evaluation} \\
\noalign{\smallskip}\svhline\noalign{\smallskip}

\citet{Ji2024} [ModelDB]&
\begin{tabular}[t]{@{}l@{}}
GPT-3.5; GPT-4\\
\end{tabular} &
API &
Author-constructed &
Accuracy, Cohen's $\kappa$, F1 \\
\hline
\citet{Dai2025} &
GPT-4o; Claude-3.5-Sonnet; Gemini-1.5-Pro &
Web UI&
Author-constructed & Recall, Specificity, AUC, SROC \\
\hline
\citet{Trad2025} &
GPT-4 &
API &
Existing syste\-matic reviews &
AER, FNR, Specificity, PPV, NPV, Efficiency \\
\hline
\citet{Hu2025} [GREP-Agent]&
GPT-4o; GPT-4o-mini; Llama-3.1; Phi-4 &
N/R &
Existing syste\-matic reviews &
Accuracy, Precision, Recall, F1 \\
\hline
\citet{Lin2025} &
Mistral-NeMo-12B &
Local &
Author-constructed &
Accuracy, Recall, Specificity, F1, Precision \\
\hline
\citet{Pan2025} &
DeepSeek-V3; Doubao-1.5-pro; Deep\-Seek-R1-Distill-Qwen-7B &
API &
Existing syste\-matic reviews &
Accuracy, Precision, Recall, F1, $\kappa$, PABAK, Efficiency, Cost \\
\hline
\citet{Mourtzinis2025} &
GPT-4.1-mini; Gemini-2.5-Flash,
Deep\-Seek-R1-70B; Llama-3.3-70B
 &
N/R &
Author-constructed &
F1 \\
\hline

\end{tabular}
}
\end{table}

\noindent\textbf{Evaluation trends.}
Regarding the ground truth, existing systematic reviews or ongoing systematic-review-style processes are used by three papers \citep{Trad2025,Hu2025,Pan2025}.  
The remaining four papers rely on author-constructed ground truth: \citet{Ji2024}'s inter-annotator-agreement gold standard, \citet{Dai2025}'s own manual screening as reference standard, \citet{Lin2025}'s single-expert review, and \citet{Mourtzinis2025}'s 20\%-sample expert-labeled subset.
Regarding the metrics used, this group again uses the standard precision/re\-call/F1/specificity core \citep{Ji2024,Dai2025,Lin2025,Hu2025,Pan2025} but several papers layer on agreement- or efficiency-oriented metrics reflecting their engineered, deployment-oriented framing. \citet{Pan2025} reports Cohen’s $\kappa$ and PABAK alongside cost and accuracy; \citet{Trad2025} use a distinctive set tailored to measuring workload reduction specifically; and \citet{Ji2024} reports Cohen’s $\kappa$ as an inter-rater agreement measure between the LLMs and experts, alongside standard accuracy and F1. \citet{Dai2025} add AUC and SROC curve analysis on top of pooled sensitivity/specificity. Finally, \citet{Mourtzinis2025} reports only F1. Columns 4-5 of Table~\ref{tab:screening_asInfra_summary} summarise the previous discussion.

\section{Discussion}
\label{sec:discussion}

Across the reviewed papers, the survey reveals a rapidly expanding body of work applying generative LLMs to both literature retrieval and screening, with the two areas showing distinct methodological profiles.
Retrieval is technically diverse, spanning semantic search, agentic query planning, knowledge-graph grounding, citation auditing, and RAG pipelines that support downstream tasks (e.g., automated review, clinical decision support, survey generation). In many systems, retrieval serves as a supporting component rather than the paper's main contribution, so authors are often more likely to construct or identify ground truth for the downstream task than for retrieval itself. As a result, focused evaluation of the LLM-based retrieval component remains uncommon.

In screening, by contrast, the methodological range is narrower, centered mainly on single-pass prompting and more engineered multi-component pipelines. Evaluation is also markedly more rigorous and consistent than in retrieval: studies are almost always benchmarked against real systematic-review decisions or carefully constructed expert-labeled ground truth, typically using an established diagnostic-accuracy toolkit. Screening therefore emerges as the more methodologically standardized of the two areas.

The survey also identifies several cross-cutting trends. First, both tasks are shifting from single LLM calls to composed, often agentic, multi-component systems. In retrieval, this appears in multi-agent pipelines for query planning, search, validation, and routing. In screening, this more often takes the form of multi-model consensus or arbitration, or escalation of uncertain cases to humans, designs that mirror how a human review team operates. This design serves not only to improve screening accuracy but also to calibrate when human involvement remains necessary. 

Second, the corpus suggests that open-weight models are gaining ground alongside still-dominant closed, API-based frontier models. OpenAI’s GPT-3.5/4/4o family remains the most common choice, with Claude and Gemini also appearing regularly. But 2025 papers increasingly use open-weight models, either alone, as in local Mistral-NeMo-12B deployment for precision-oncology curation, or alongside closed models in mixed pipelines. Where authors give reasons, the main drivers are privacy-preserving local deployment and cost control in workflows requiring many repeated or parallel calls. Direct open-versus-closed comparisons, however, remain too limited for firm conclusions.

Third, access and adaptation are unevenly reported in both areas. API access to commercial frontier models dominates, but retrieval papers often leave the access mode unstated, whereas it is confirmed for nearly every screening paper. At the same time, domain-specific fine-tuning is rare.

Fourth, the corpus is heavily concentrated in biomedical and clinical applications. The most common topics are biomedical (see Section~\ref{sec:corpus:stats}), and most presented engineered screening pipelines, along with several RAG-based retrieval systems, target medical evidence synthesis or clinical decision support. This likely reflects both the maturity of systematic-review methodology in biomedicine, which provides established ground truth and review conventions, and the field's strong need for scalable evidence synthesis. As a result, the patterns identified here are best understood as characteristic primarily of LLM-assisted biomedical literature work and may not transfer directly to domains without similar benchmarks or review practices.

\section{Conclusion}
\label{sec:conclusion}

This chapter has surveyed $34$ peer-reviewed papers published between 2024 and 2026 that apply generative LLMs to two core scientific knowledge-discovery tasks, literature retrieval and literature screening, motivated by the growing strain that expanding publication volumes place on traditional, largely manual search and review workflows.

The picture that emerges is of an active and rapidly growing but highly concentrated body of work. The great majority of the reviewed systems target biomedical and clinical applications, particularly systematic reviews and other evidence-synthesis workflows. This is unsurprising since biomedicine offers mature review methodology, ready-made ground truth, and clear evaluation conventions, while also facing a particularly acute need to keep pace with a fast-growing evidence base. Outside this setting, comparable benchmarks and conventions are largely absent, so the transferability of these approaches to other domains remains unclear.

Taken together, the surveyed work suggests that LLM-assisted literature retrieval and screening is moving from isolated demonstrations toward a more established, if still domain-bound, area of practice.

\bibliographystyle{spbasic}
\bibliography{main, references}

\end{document}